\documentclass[conference]{IEEEtran}
\IEEEoverridecommandlockouts
\usepackage{cite}
\usepackage{amsmath,amssymb,amsfonts}
\usepackage{algorithmic}
\usepackage{graphicx}
\usepackage{textcomp}
\usepackage{multirow}
\usepackage{makecell}
\usepackage{subcaption}
\usepackage{xcolor}
\usepackage{booktabs} 
\usepackage{listings} 

\def\BibTeX{{\rm B\kern-.05em{\sc i\kern-.025em b}\kern-.08em
    T\kern-.1667em\lower.7ex\hbox{E}\kern-.125emX}}
\begin{document}

\title{Advanced Learning-Based Inter Prediction for Future Video Coding \\
\thanks{$^{\ast}$Corresponding author: Siwei Ma. This work was supported in part by the National Natural Science Foundation of China under grant no. 62025101 and 62031013, and in part by the High Performance Computing Platform of Peking University, which are gratefully acknowledged.}
}

\author{
\IEEEauthorblockN{Yanchen Zhao$^{1}$, Wenhong Duan$^{1}$, Chuanmin Jia$^{2}$, Shanshe Wang$^{1}$ and Siwei Ma$^{1,\ast}$}
\IEEEauthorblockA{$^{1}$School of Computer Science, Peking University, China
}
\IEEEauthorblockA{$^{2}$Wangxuan Institute of Computer Technology, Peking University, China
}
\IEEEauthorblockA{yczhao@stu.pku.edu.cn, whduan@sjtu.edu.cn, \{cmjia, sswang, swma\}@pku.edu.cn
}
}



\maketitle

\begin{abstract}
In the fourth generation Audio Video coding Standard (AVS4), the Inter Prediction Filter (INTERPF) reduces discontinuities between prediction and adjacent reconstructed pixels in inter prediction. The paper proposes a low complexity learning-based inter prediction (LLIP) method to replace the traditional INTERPF. LLIP enhances the filtering process by leveraging a lightweight neural network model, where parameters can be exported for efficient inference. Specifically, we extract pixels and coordinates utilized by the traditional INTERPF to form the training dataset. Subsequently, we export the weights and biases of the trained neural network model and implement the inference process without any third-party dependency, enabling seamless integration into video codec without relying on Libtorch, thus achieving faster inference speed. Ultimately, we replace the traditional handcraft filtering parameters in INTERPF with the learned optimal filtering parameters. This practical solution makes the combination of deep learning encoding tools with traditional video encoding schemes more efficient. Experimental results show that our approach achieves 0.01\%, 0.31\%, and 0.25\% coding gain for the Y, U, and V components under the random access (RA) configuration on average.
\end{abstract}


\section{Introduction}
In recent years, video has gradually become the primary medium for information dissemination. The exponential growth of video data has created new challenges for image and video compression. Over the past several decades, numerous influential standards and specifications have been developed to address these challenges, such as ISO/IEC MPEG series~\cite{tudor1995mpeg,haskell1996digital,sikora1997mpeg,li2001overview}, ITU-T H.26X series~\cite{sullivan2004h,vetro2011overview,sullivan2012overview,bross2021overview}, Audio Video coding Standard (AVS) series~\cite{he2013framework,zhang2019recent} and the AV1~\cite{chen2018overview,han2021technical} from the Alliance of Open Media (AOM). These frameworks focus on prediction coding~\cite{gao2020advanced}, transformation and quantization coding~\cite{quantization2019}, entropy coding~\cite{sze2012entropy}, as well as in-loop filtering~\cite{Tsai2013Adaptive} to efficiently reduce spatial, temporal and statistical redundancies. Recent advancements in video coding standards, such as Versatile Video Coding (VVC) and AVS3, emphasize enhancing the adaptability and versatility of video codecs to cater to diverse application scenarios. These latest standards are designed to deliver higher compression efficiency and improve quality while accommodating a wide range of content and devices.

\begin{figure}[tbp]
    \centering
    \includegraphics[scale=1]{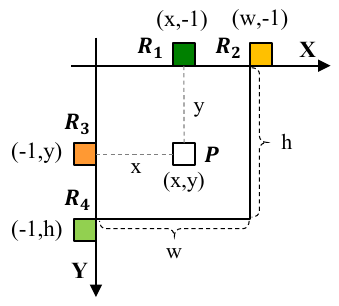}
    \caption{Inter Prediction Filter (INTERPF) in Audio Video coding Standard (AVS) traverses each pixel $P$ in the current Coding Unit (CU) block and then filters pixel $P$ using its coordinate values and adjacent four reconstructed pixel values.}
    \label{fig:interpf}
    \vspace{-0.5cm}
\end{figure}

However, the traditional hybrid coding framework is still struggling to efficiently compress ultra-high definition video. With the success of deep learning in image and video processing, deep learning for video coding~\cite{ma2019image,duan2023learned} is an emerging area of research. In this context, ad hoc groups (AHGs) have been established in AVS and the Joint Video Experts Team (JVET) to explore video coding tools based on deep learning. Various neural network-based coding tools have been proven effective in integrating with video codecs, including neural network-based intra prediction~\cite{dumas2019context,dumas2021combined}, deep reference frame techniques~\cite{zhao2019vrf,10201390}, in-loop filter~\cite{10.1145/3529107}, and frame super-resolution~\cite{super-resolution,10533741}. These deep learning-based encoding tools have demonstrated remarkable improvements in compression performance and reconstruction quality, making them crucial in advancing intelligent video coding standards. However, the computational complexity of neural network-based coding tools is much higher than that of traditional coding tools, which puts a heavy burden on the standardization and application of neural network tools.

\begin{figure*}[tbp]
    \centering
    \includegraphics[width=1\textwidth]{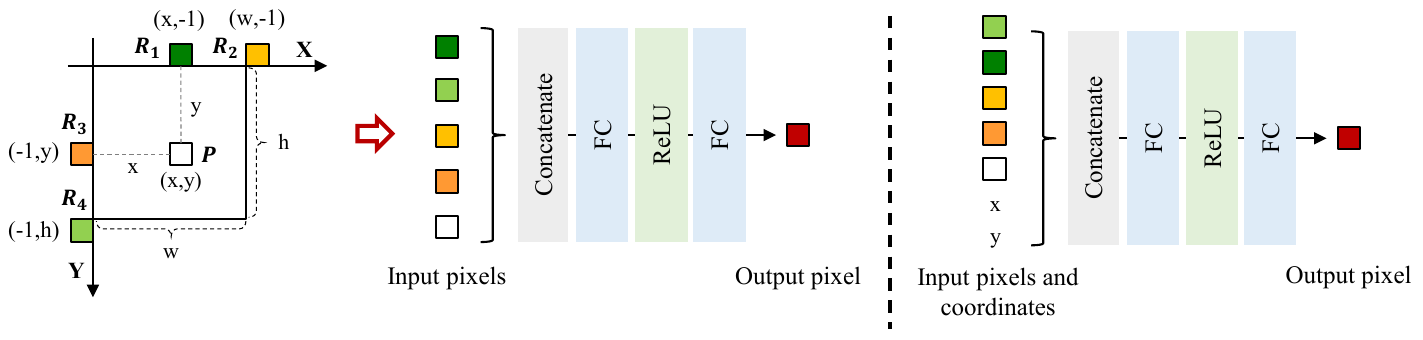}
    \caption{Illustration of the low complexity learning-based inter
prediction network architecture. The figure shows two different schemes, with ``Scheme 1'' on the left, which only inputs the values of the adjacent reconstructed pixels and the current pixel. On the right is ``Scheme 2'', which adds the coordinates of the current pixel in addition to ``Scheme 1''.}
    \label{fig:network}
    \vspace{-0.5cm}
\end{figure*}

In this paper, different from previous neural network-based coding tools for hybrid coding frameworks with module-level substitution and the addition of decision-making modes, we propose a learning-based optimization method for traditional statistical prior-based coding tools. We conduct experiments on the traditional Inter Prediction Filter (INTERPF)~\cite{interpf}. The contributions can be summarized as follows:
\begin{itemize}
    \item We propose a lightweight fully connected network that can replace the traditional INTERPF and achieve significant coding gains while maintaining complexity comparable to traditional coding tools.
    \item We propose an inference library implemented without any third-party dependency. This is more conducive to combining neural network-based coding tools with traditional codecs and has a faster inference speed compared to using the Libtorch\footnote{https://pytorch.org/} library. 
    \item We propose an effective solution to use neural networks to optimize traditional coding tools, which provides a new direction for exploring the next generation of video coding tools by leveraging lightweight models and an efficient inference library.
\end{itemize}

\section{Inter Prediction Filter in AVS}
In Audio Video coding Standard (AVS), the Inter Prediction Filter (INTERPF) refines the inter prediction blocks obtained from the direct prediction mode to eliminate the discontinuities between the prediction pixels and the adjacent reconstructed pixels. As shown in Fig. \ref{fig:interpf}, it uses four adjacent reconstructed pixels $R_1$, $R_2$, $R_3$, $R_4$ and the relative distance between the current prediction pixel $P$ and the reconstructed pixels to refine the current prediction pixel $P$.

\begin{equation}
    \begin{split}
P_V(x,y) =& ((h-1-y)*R_1 + (y+1)*R_4 \\
&+(h>>1))>>\log_2(h),
\label{eq1} 
    \end{split}
\end{equation}

\begin{equation}
    \begin{split}
P_H(x,y) =& ((w-1-x)*R_3 + (x+1)*R_2\\
&+ (w>>1))>>\log_2(w),
\label{eq2} 
    \end{split}
\end{equation}

\begin{equation}
P_Q(x,y) = ( P_V(x,y) + P_H(x,y) + 1 ) >> 1,
\label{eq3}
\end{equation}

\begin{equation}
O(x,y) = ( P(x,y) * 5 + P_Q(x,y) * 3 + 4 ) >> 3.
\label{eq4}
\end{equation}

Specifically, as shown in Eq. \ref{eq1}, Eq. \ref{eq2}, Eq. \ref{eq3}, and Eq. \ref{eq4}, the inter prediction filter uses a weighted average method to obtain the final optimized prediction value. To control the complexity of encoding and decoding, the tool is only enabled in the direct inter prediction mode. This method also requires adding an inter prediction filtering flag in the bitstream, indicating whether the current coding block enables inter prediction filter. This method was tested on the AVS reference software High-Performance Model (HPM), under the Random Access (RA) configuration. It can achieve 0.34\%, 0.04\%, and 0.13\% coding gains on the Y, U, and V components, respectively. Due to the limitations of handcrafted algorithms based on statistical priors, which cannot adapt to diverse video content, the improvement in coding gain is limited.

\section{Proposed Method}
\subsection{Network Structure Analysis}
The network architecture is constructed by two fully connected layers. Rectified Linear Unit (ReLU) is used as the activation function after the first fully connected layer. As shown in Fig. \ref{fig:network}, we design two different experimental configurations. The first configuration takes the prediction value of the current pixel $P$ at $(x, y)$ and the $(x, -1)$, $(w, -1)$, $(-1, y)$, $(-1, h)$ reconstructed values of the four adjacent pixels $R_1$, $R_2$, $R_3$, $R_4$ as the input $I_1$, as given in Eq. \ref{equ:inputsimple}. As shown in Eq. \ref{equ:inputimprove}, the second configuration is to additionally input the coordinate values $x$, $y$ of the current pixel $P$ in addition to the input $I_1$. The output of the network is the filtered prediction value of the current pixel $P$. Tab. \ref{tab:netstruct} shows the structure of each layer of the network and the size of input and output. We align the number of neurons in the first layer of the network with the size of the input to learn the correlation between different inputs. The final layer of the network fuses the outputs of the hidden layers to obtain the refined prediction pixel. 

\begin{equation}
\label{equ:inputsimple}
    I_1=(R_1,R_2,R_3,R_4,P),
\end{equation}

\begin{equation}
\label{equ:inputimprove}
    I_2=(R_1,R_2,R_3,R_4,R_5,P,x,y).
\end{equation}

\begin{table}[t]  
\centering
\caption{The network architecture has 2 fully connected (FC) layers. The input FC layer has 5 and 7 neurons in the two schemes respectively. The output FC layer has 1 neuron.}
\resizebox{1\linewidth}{!}{
\begin{tabular}{ccc}
\bottomrule[2pt]
\textbf{} & \textbf{Scheme 1} & \textbf{Scheme 2}\\
\hline
Input:& Input dimension ($1 \times 5$) & Input dimension ($1 \times 7$)\\
\hline
[layer 1]& FC (5 neurons) & FC (7 neurons) \\

[layer 2]& ReLU & ReLU \\

[layer 3]& FC (1 neuron) & FC (1 neuron) \\
\hline
Output:& Output dimension ($1 \times 1$) & Output dimension ($1 \times 1$) \\
\toprule[2pt]
\end{tabular}
}
\label{tab:netstruct}
\vspace{-0.5cm}
\end{table}

\subsection{Model Inference Library}
To accelerate the model inference stage and facilitate application deployment, we implement the fully connected network inference process using pure C++ and propose a neural network model inference library for video coding without third-party dependency. Specifically, we export the weights and biases for each layer in the model as single-precision floating-point values. We preload these weights and biases into the global float variables when the codec is initialized. During the inference stage, the input values, weights, and biases are multiplied and added through the CPU. This approach allows us to fully disengage from the Libtorch library, simplifying code execution and deployment while improving inference speed compared to using the Libtorch library. 

For the first configuration, the fully connected network has 5 neurons in the first layer and 1 neuron in the last layer, with an input size of 5. We export 30 weights and 6 biases, resulting in a total of 36 values. For the second configuration, the fully connected network has 7 neurons in the first layer and 1 neuron in the last layer, with an input size of 7. We export 56 weights and 8 biases, totaling 64 values.

\subsection{Implementation in EVM}
We conduct experiments on the next-generation video coding platform Exploration Video Model (EVM) of Audio Video coding Standard (AVS). For code implementation in EVM, there is no change to the traditional inter prediction filter (INTERPF) filtering process, only replacing the computation of $O(x, y)$ in Eq. \ref{eq4}. As shown in Eq. \ref{eqmodel}, currently $O(x, y)$ is calculated through the trained models,
\begin{equation}
O(x,y) = f(I;\theta),
\label{eqmodel}
\end{equation}
where $O(x,y)$ represents the filtered prediction value of the current pixel, $f$ represents the filtering task, $I$ represents the input values, and $\theta$ represents the set of learnable parameters in the network.

\section{Experimental Results}
\subsection{Traning Process}
The training is a regression problem from adjacent reconstructed pixels $R_i$ and current pixel $P$ to the original pixel $G$ with the network parameter $\theta$. The loss function is mean-square error (MSE) as shown in Eq. \ref{equ:mse},
\begin{equation}
\label{equ:mse}
J(\theta)=\frac{1}{N} \sum_{i=0}^{N-1}\left \|f(I^{i};\theta)-G^{i} \right \|_{2},
\end{equation}
where the batch size $N$ is 1000. We use Class B, C, and D of the BVI-DVC dataset~\cite{ma2021bvi} with a total of 600 videos as the training set. We convert the original MP4 format videos into 10-bit YUV420 format videos by ffmpeg\footnote{https://ffmpeg.org/}. Then we encode the first 32 frames of each video with four quantization parameters (QPs) (27, 32, 38, 45) by the Audio Video Coding Standard (AVS) latest reference software Exploration Video Model (EVM) under random access (RA) configuration. Other parameters are the same as AVS common test conditions (CTC)~\cite{EVMctc}. We modify the code of the INTERPF tool on the decoding side to be able to export all pixels of all coding units (CUs) filtered using the traditional INTERPF tool, as well as their adjacent reconstructed pixels. To pair with the original pixels, we also export the coordinates, width, height, and picture order count (POC) of each CU. Adam~\cite{kingma2014adam} is used as the optimizer. The initial learning rate is set to 1e-4 and reduced to 3e-5 at the last epoch. More detailed training information is shown in Tab. \ref{tab:trainingandinfer}.

In addition, for ``Scheme 1'', we train a single set of model parameters using a training set comprising Class B, Class C, and Class D videos. In contrast, ``Scheme 2'' involves training six distinct sets of model parameters, differentiated by Y, U, and V components and video resolutions. Specifically, we train three sets of parameters for the Y, U, and V components using a training set from Class B for inter prediction in 1080p and 4K videos. Additionally, we trained another three sets of parameters for the Y, U, and V components using a training set from Class C and Class D for inter prediction in 720p videos.

\begin{table}[tbp]  
\centering
\caption{Training stage and inference stage information.}
\resizebox{1\linewidth}{!}{
\begin{tabular}{c|c}
\bottomrule[2pt]
\multicolumn{2}{c}{\textbf{Training Stage}}\\
\hline
\hline
\textbf{GPU Type:}&NVIDIA RTX 4090D 24GB\\
\hline
\textbf{CPU Type:}&Intel i9-14900K\\
\hline
\textbf{Framework:}&PyTorch\\
\hline
\textbf{Epoch:}&3\\
\hline
\textbf{Batch Size:}&1000\\
\hline
\textbf{Training Time:}&9 hours\\
\hline
\textbf{Loss Function:}&MSE\\
\hline
\textbf{Training Sequences:}&BVI-DVC (Class B, Class C, Class D)\\
\hline
\textbf{Learning Rate:}&1e-4, 3e-5\\
\hline
\textbf{Optimizer:}&Adam\\
\hline
\bottomrule[2pt]
\multicolumn{2}{c}{\textbf{Inference Stage}}\\
\hline
\hline
\textbf{CPU Type:}&Intel Xeon Platinum 8358 \\
\hline
\textbf{Framework:}&Proposed inference library\\
\hline
\textbf{Complexity:}&Scheme 1: 30 MACs; Scheme 2: 56 MACs\\
\hline
\textbf{Number of Models:}&Scheme 1: 3; Scheme 2: 6\\
\hline
\textbf{Total Parameters Number:}& Scheme 1: $108$; Scheme 2: $384$\\
\toprule[2pt]
\end{tabular}
}
\label{tab:trainingandinfer}
\vspace{-0.5cm}
\end{table}

\subsection{Coding Gain Analysis}
The proposed method is implemented into EVM-0.4~\cite{EVM0.4}. We strictly follow the coding configuration of RA given by AVS CTC. Four QPs (27, 32, 38, 45) are tested, and the coding efficiency comparison with the anchor (EVM-0.4) is measured by BD-rate. First, we test ``Scheme 1'' by inputting only the adjacent reconstructed pixels and the current prediction pixel. Tab. \ref{tab:5inputvsEVM0.4interpfon} summarizes the average BD-rate savings under RA configuration. The results show that with only pixel values and no coordinates, it only achieves comparable performance to the traditional INTERPF, and no further performance gains can be achieved. This is mainly due to the difficulty for the model to learn generalized filtering parameters from the data exported from videos with different CU sizes and resolutions in the training set. Due to the small number of parameters in this model and the different characteristics of Y, U, and V components, it is difficult to learn a generalized set of filtering parameters that can simultaneously ensure rate-distortion performance in both luminance and chrominance components.

Then we test the ``Scheme 2''. Tab. \ref{tab:7inputvsEVM0.4interpfon} shows the performance comparison with the EVM-0.4 when inputting additional coordinates. The proposed method can achieve 0.01\%, 0.31\%, and 0.25\% BD-rate savings on average under RA configuration. Adding the coordinates of the current pixel and training the model separately for different resolutions, luminance, and chrominance components can effectively improve the coding gain.

\begin{table}[tbp]
\caption{Performance of EVM-0.4 combined with the ``Scheme 1'' under RA configuration compared with the EVM-0.4.}
\vspace{-0.3cm}
\begin{center}
\setlength{\tabcolsep}{1pt}
\renewcommand{\arraystretch}{1.1}
\resizebox{1\linewidth}{!}{
\begin{tabular}{c|c|ccc|c|c}
\bottomrule[2pt]
\multirow{2}{*}{\textbf{Class}} & \multirow{2}{*}{\textbf{Sequence}} & \multicolumn{5}{c}{\textbf{Random Access}} \\
\cline{3-7}
                       &                          & \textbf{Y} & \textbf{U} & \textbf{V} & \textbf{EncT} & \textbf{DecT}\\
\hline
\multirow{4}{*}{UHD4K}       &Tango2&0.08\%&0.21\%&-0.06\% &\multirow{4}{*}{104\%}&\multirow{4}{*}{104\%} \\
                             &ParkRunning3&0.02\%&0.00\%&0.03\%&&\\
                             &Campfire&0.04\%&0.08\%&-0.13\%&&\\
                             &DaylightRoad2&-0.04\%&-0.02\%&-0.21\%&&\\
\hline
\multirow{4}{*}{1080p}       &Cactus
                             &0.04\%&0.29\%&-0.17\%&\multirow{4}{*}{102\%}&\multirow{4}{*}{103\%} \\
                             &BasketballDrive  &0.16\%&0.33\%&-0.21\%&&\\
                             &MarketPlace  &0.03\%&0.06\%&-0.13\%&&\\
                             &RitualDance  &0.06\%&0.47\%&0.40\%&&\\
                    
\hline
\multirow{4}{*}{720p}        &City  
                             &-0.02\%&-0.24\%&-0.11\%&\multirow{4}{*}{104\%}&\multirow{4}{*}{104\%} \\
                             &Crew  &0.09\%&-0.08\%&0.35\%&&\\
                             &Vidyo1  &-0.15\%&-0.14\%&-0.11\%&&\\
                             &Vidyo3  &-0.06\%&-0.37\%&-0.10\%&&\\
\hline
\multicolumn{2}{c|}{\textbf{Average}}  &\textbf{0.02\%}&\textbf{0.05\%}&\textbf{-0.04\%}&\textbf{103\%}&\textbf{104\%}\\
\toprule[2pt]
\end{tabular}
}
\label{tab:5inputvsEVM0.4interpfon}
\end{center}
\vspace{-0.7cm}
\end{table}

\subsection{Coding Complexity Analysis}
To evaluate the performance of different inference methods in the video encoding and decoding process, we conduct experiments on the ``BasketballPass\_416x240\_50.yuv'' video sequence. We compare the coding time required by two different approaches: using the Libtorch library and our custom inference library. Experiments are conducted on the following hardware specifications: an Intel Core i7-12700 CPU and a Nvidia GeForce RTX 3060 Ti GPU. We conduct our experiments using the RA configuration, encoding only the first 32 frames of each video. To ensure fairness, we employ a model preloading strategy in both implementations, eliminating any extra time due to repeated model loading. Tab. \ref{tab:inferstagetime} shows the coding complexity comparison between the two approaches. During the encoding process, inference with the Libtorch library takes 5268 seconds in total, averaging 164.6 seconds per frame. In contrast, our method completes the encoding process in just 310 seconds in total, with an average of 9.7 seconds per frame, achieving a $17\times$ encoding speedup and a $3\times$ decoding speedup. For our encoding and decoding tests, we use the following command line instructions:

\begin{itemize}
\item encoder\_app.exe --config encode\_RA.cfg -i BasketballPass\_416x240\_50.yuv -w 416 -h 240 -z 50 -d 10 -f 32 -q 45 -o BasketballPass\_416x240\_50.bin
\item decoder\_app.exe -i BasketballPass\_416x240\_50.bin
\end{itemize}

\begin{table}[tbp]
\caption{Performance of EVM-0.4 combined with the ``Scheme 2'' under RA configuration compared with the EVM-0.4.}
\vspace{-0.3cm}
\begin{center}
\setlength{\tabcolsep}{1pt}
\renewcommand{\arraystretch}{1.1}
\resizebox{1\linewidth}{!}{
\begin{tabular}{c|c|ccc|c|c}
\bottomrule[2pt]
\multirow{2}{*}{\textbf{Class}} & \multirow{2}{*}{\textbf{Sequence}} & \multicolumn{5}{c}{\textbf{Random Access}} \\
\cline{3-7}
                       &                          & \textbf{Y} & \textbf{U} & \textbf{V} & \textbf{EncT} & \textbf{DecT}\\
\hline
\multirow{4}{*}{UHD4K}       &Tango2&-0.03\%&0.13\%&-0.03\% &\multirow{4}{*}
                            {102\%}&\multirow{4}{*}{101\%} \\
                             &ParkRunning3&0.05\%&0.04\%&-0.01\%&&\\
                             &Campfire&0.08\%&-0.03\%&-0.11\%&&\\
                             &DaylightRoad2&0.06\%&-0.34\%&-0.09\%&&\\
\hline
\multirow{4}{*}{1080p}       &Cactus
                             &-0.03\%&-0.84\%&-0.57\%&\multirow{4}{*}
                             {105\%}&\multirow{4}{*}{108\%} \\
                             &BasketballDrive  &0.00\%&0.12\%&-0.10\%&&\\
                             &MarketPlace  &-0.01\%&-0.40\%&-0.23\%&&\\
                             &RitualDance  &-0.02\%&-0.16\%&0.14\%&&\\
                    
\hline
\multirow{4}{*}{720p}        &City  
                             &-0.03\%&-0.90\%&0.37\%&\multirow{4}{*}{104\%}&\multirow{4}{*}{105\%} \\
                             &Crew  &-0.13\%&1.13\%&0.18\%&&\\
                             &Vidyo1  &-0.06\%&-0.83\%&-0.28\%&&\\
                             &Vidyo3  &-0.02\%&-1.63\%&-2.24\%&&\\
\hline
\multicolumn{2}{c|}{\textbf{Average}}  &\textbf{-0.01\%}&\textbf{-0.31\%}&\textbf{-0.25\%}&\textbf{103\%}&\textbf{105\%}\\
\toprule[2pt]
\end{tabular}
}
\label{tab:7inputvsEVM0.4interpfon}
\end{center}
\vspace{-0.4cm}
\end{table}

\begin{table}[t]
\caption{Comparison of coding complexity.}
\centering
\resizebox{1\linewidth}{!}{
\begin{tabular}{c|c|c}
\bottomrule[2pt]
\textbf{Inference Libarary} & \textbf{Encoding Time} & \textbf{Decoding Time}\\
\hline
Libtorch (with GPU)& 5268.544 seconds & 0.324 seconds\\
\hline
Proposed Library (CPU only)& 310.869 seconds & 0.098 seconds\\
\hline
Anchor (CPU only)& 299.893 seconds & 0.094 seconds \\
\toprule[2pt]
\end{tabular}
}
\label{tab:inferstagetime}
\vspace{-0.5cm}
\end{table}
$i$ represents the file name of the input video or bitstream. $o$ represents the file name of the output bitstream. $w$ and $h$ represent the width and height of the input video. $z$ represents the frame rate. $d$ represents the bit depth of the input video. $f$ represents the maximum number of frames to be encoded. $q$ represents the QP value. Other parameters are set according to the configuration file ``encode\_RA.cfg''.

Note that, during the inference stage with Libtorch, the primary source of time consumption is the transfer of data between system memory and graphics memory, rather than the model inference. This data transfer latency is an obvious bottleneck that hampers the efficient integration of deep learning-based coding tools with traditional codecs.

\section{Conclusion}
In this paper, we proposed a low complexity learning-based inter prediction (LLIP) method. By exporting the model's weights and biases and implementing the inference process without any third-party dependency, we can achieve better rate-distortion performance with minimal computational overhead. This approach reduced complexity effectively and can provide a new direction for exploring the next generation of video coding tools by leveraging lightweight models and efficient inference techniques.




\bibliographystyle{IEEEtran}
\bibliography{ref}{}

\end{document}